# Detectors for Time-of-Flight Fast-Neutron Radiography
## 1. Neutron Counting Gas Detector

V. Dangendorf*, G. Laczko, M. Reginatto

Physikalisch Technische Bundesanstalt (PTB), Braunschweig, Germany

D. Vartsky, M. Goldberg, I. Mor

*Soreq NRC, Yavne, Israel*

A. Breskin, R. Chechik

*Weizmann Institute of Science, Rehovot, Israel*

**Abstract**

One of our two methods for fast-neutron imaging with spectrometric capability is presented here. It is a neutron-counting technique based on a hydrogenous neutron converter coupled to Gaseous Electron Multipliers (GEM). The principles of the detection techniques and the optimization of the converter, electron amplification and the readout are described. Evaluation of the properties are derived from a experiment in a pulsed neutron beam of spectral distribution between 2 and 10 MeV.

### 1. Introduction

Recently we have reported about the development of two imaging methods for Fast-Neutron Resonance Transmission Radiography (FNRT) [1,2]. In our approach to this technology a pulsed, broad energy neutron beam is produced by a nanosecond pulsed deuterium beam hitting a thick Be target. Selected neutron-energy windows, relevant for element identification in a sample, are selected by Time-Of-Flight (TOF) techniques. The most challenging feature in this approach is the requirement of efficient, large area neutron imaging detectors with nanosecond timing capability. In our earlier articles we have proposed and demonstrated the capabilities of different detector concepts. One is a pulse-counting detector based on a hydrogenous neutron converter coupled to a position-sensitive gas detector. The second one is based on a fast plastic scintillator screen viewed by a nanosecond-gated fast framing camera or several independently gated intensified CCD cameras. The third method uses the scintillator based optical system but the gated intensifier is replaced by a pulse counting optical imaging detector with sub-nanosecond timing capability. All these systems are able to capture images with one millimetre (and better) spatial resolution and < 10 ns TOF resolution.

* Corresponding author. , Tel: +49 531 592 7525; fax: +49 531 592 7205; e-mail: volker.dangendorf@ptb.de

The earlier status in the development of these detectors and our neutron facility is presented in [1]. A short report on recent developments of the imaging systems can be found in [2]. This paper focuses on the description of the pulse counting gas detector (FANGAS, from FAst Neutron GASeous imaging detector. We describe the concept of the detection method and the technical realisation of one module. Experimental results obtained in neutron beam experiments at the PTB neutron facility are presented. A detailed description and evaluation of the optical systems will be published elsewhere.

## 2. The Detector

### 2.1 The FANGAS concept

The first step in neutron detection is the efficient conversion of neutrons into energetic charged particles which can be detected by established methods due to their light or electric charge carrier production. In earlier work we have reported on the development of a thermal neutron imaging methods based on solid foil Gd and $^{6}$Li neutron converters [3-6]. The method is based on detecting electrons or ions from neutron interactions in a converter foil. These particles can escape into a gas and are detected by one or several position sensitive wire chambers. With this method thermal neutron detectors with about 25 % detection efficiency, sub-millimeter position resolution and about 100 ns time resolution were realized. In FANGAS, this method is developed further to satisfy the needs for application in fast-neutron resonance radiography in pulsed beams. The goals for the detectors in this application can be summarized as follows:

- Detection area: > 30x30 cm$^2$
- Detection efficiency: > 5 %
- Insensitivity to gamma-rays
- Count rate capability > 10$^6$
- Spectroscopic properties in 2-10 MeV range
- Energy resolution: < 500 keV at 8 MeV
- Position resolution (fwhm): 0,5 mm

Because fast neutron cross section are only of the order of a few bn (compared to thousands to hundreds of thousand kbn at thermal energies), the conversion efficiency of the best neutron to charge particle converters for MeV neutrons are much smaller. Therefore, to achieve detection efficiencies of 5 %, individual detectors modules have to be cascaded along the beam axis. This was already proposed for the thermal neutron detector based on a $^{6}$Li converter foil to achieve higher efficiencies [6]. For fast neutrons about 25 modules are needed, which necessitates to develop techniques for simple and rugged construction and handling, to minimize production cost and time of trained person involvement during operation.

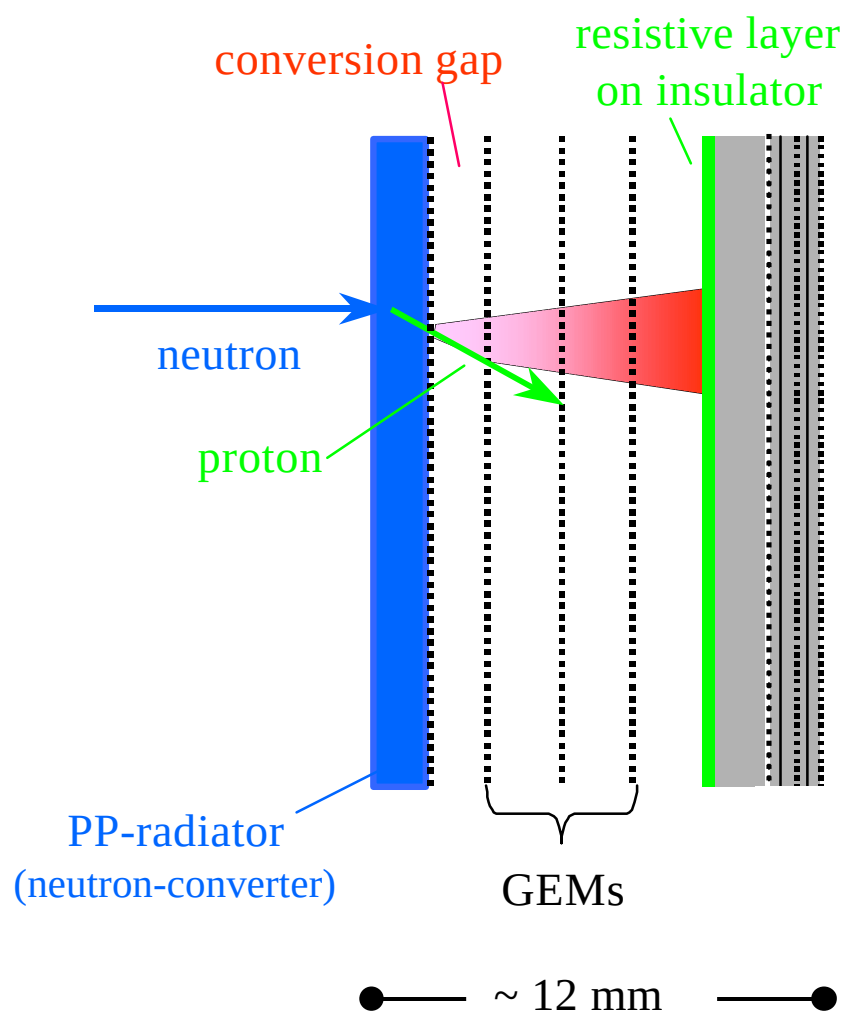

Fig. 1: Schematic drawing of the fast-neutron gaseous detector

### 2.2 The detection method

Fig. 1 shows the principle of the detection method. In the converter stage neutrons scatter with protons in a hydrogen rich radiator foil, made of polypropylene (PP) or polyethylene (PE). Protons can escape from the foil into an the adjacent 0,5 - 1 mm wide gas filled conversion gap of an electron multiplier. Here the protons produce ionisation electrons which are collected and multiplied by 3 cascaded Gas Electron Multipliers (GEMs) [7]. After leaving the last GEM, the amplified electron cloud travels through the induction gap and is finally collected on an anode. This anode is a thin film of high surface resistivity, deposited on an insulating substrate. Through this layer the moving charge in the induction gap induces a signal on a readout (RO) board behind the insulating substrate. The readout board is a double layer printed circuit board (PCB) which contains on both sides a pad structure for position encoding. The position read out is obtained by the delay line technique.

**2.3 The proton radiator**

The neutron radiator is optimized to satisfy 2 conflicting requirements:

a) neutrons have to interact with radiator material and produce energetic charged particles. This emphasizes thick converters with large neutron cross sections.

b) the charged particles must reach the surface of the foil and escape into the gas volume. This favors thin foils as well as reactions where the secondary particles are of low Z (e.g. protons) and where these particles receive high kinetic energy. Furthermore, materials of low stopping power ($dE/dx$) are preferable.

The best radiator candidates are hydrocarbon foils, where the neutron scatters elastically with hydrogen. PE and PP were chosen because they are easy to use and have the largest hydrogen content of all plastic foils. Since the converter in the detector has also to act as electrode the converter was coated with a thin layer of graphite.

The parameters of the neutron radiator and the ion converter are estimated and optimized by simulating the neutronics and the charged particle transport with GEANT [8]. The details and the results of these simulations are published elsewhere in this proceedings [9]. For our detector, designed for energies between 2 and 10 MeV, a 1 mm converter thickness was chosen which provides for 2 MeV and 7,5 MeV neutrons a detection efficiency of 0,05 % and 0,2 % respectively.

The physical limit of the position resolution, defined by the point spread function (PSF), is also calculated in [9]. Protons escape from the radiator surface into the conversion gap at various angles. The detected position of these protons is assumed to be in the middle of the proton track inside the1 mm wide conversion gap. The distribution of these positions for a neutron beam of infinitesimal small size is assumed to define the ideal PSF. For the relevant energy range from 2 – 10 MeV a fwhm resolution of 0,5 - 1 mm can be achieved. Because this is the physical limit, the subsequent charged particle detection and read out techniques have to provide only this level of accuracy and can be designed accordingly.

**2.4. Ion conversion and electron amplification**

The proton, after escaping form the radiator, enters a 1 mm wide gas filled conversion gap where it produces electrons by ionising the gas molecules. For convenience we used a standard gas mixture for GEM operation, Ar (70 %) and $CO_2$ (30%) at a pressure of 1 bar. The number or electrons produced by protons in the relevant energy range of 100 keV – 10 MeV ranges from about 200 to 4000 in a 1 mm gap (energy loss calculated by SRIM[10]). The width of the conversion gap is a compromise between the need of a thin gap for obtaining the highest position resolution and the difficulty of manufacturing thin parallel gaps in large area detectors. The ionisation electrons are efficiently amplified by 3 cascaded GEMs made by the CERN group [7]. For the prototype module we have used 10 by 10 cm square GEMs on 70 µm thick Kapton, with ∅80 µm holes and 140 µm pitch. The detector electrodes are biased via a resistor network with equal voltage steps on all GEMs and in between them. Sufficient gain is obtained at about 405 V on each GEM. The voltage in the conversion gap is individually adjustable to optimize for maximum collection efficiency (at about 150 V). The voltage across the induction gap (between the last GEM and the collection anode) is raised to twice the value of the GEM voltage to ensure efficient charge transfer and rapid charge collection on the anode. The width of the induction gap was varied between 1 and 3 mm and the influence on the signal shape and amplitude at the pickup electrodes was studied (see also [11]). The best results were obtained at 1 mm where signal duration matched the time constant of the delay lines and preamplifiers, and the amplitude was at its maximum.

**2.5. Resistive anode and signal induction on redout electrode**

For position readout, the method of induced signal readout (RO) through a resistive anode was chosen [12-16]. This method has the following advantages compared to the direct readout on a pad or strip structured anode:

a) spread of the geometrical size of the induced signal:
the small radial size of the final avalanche cloud is frequently a problem when using GEMs at atmospheric pressure. Applying an interpolating RO method (delayline, wedge-and-strip...) the readout structures have to be either very small (of the order of 100 μm) or the induction gap must be sufficiently wide to enable radial diffusion of the avalanche. The induced charge method allows the use of a 2 mm pitch on the RO structure with a only 1 mm wide induction gap.
b) galvanic decoupling of detector and RO board:
The resistive anode can be operated at high voltage while the RO board is at ground potential.
c) protection of RO circuits:
electrical discharges in the detector do not damage the sensitive RO electronics because the electromagnetic shock is limited by the small serial capacitance between anode and RO board.
To maximize the signal amplitude on the RO we have investigated several resistive anode coatings. The requirements are:

a) maximum signal on the RO boards
b) inexpensive material and easy deposition
c) lowest possible surface resistance in order not to disturb detector operation at high count rate by changes in the anode potential.

Vacuum deposited Ge layers and sprayed graphite paints [16] of different thicknesses were investigated. A signal transmission of 100 % was achieved with Ge at surface resistivities of 30 – 500 MΩ . The sprayed anodes were of 1 – 5 MΩ and their transmission ranged from 50 – 85 %. Due to their simple deposition technique and their proven stability during operation the graphite anodes are the first choice for a future multi-element detector, despite their lower signal transparency.

## 2.6. The read out electrode

The position signals are derived from a pad-structured RO electrode which was proposed by Jagutzki et al.[12]. Fig 2 shows a schematic drawing of the RO board. Pads for X and Y coordinates are printed on both sides of a standard, 0,5 mm thick PCB and are interconnected by thin strips in orthogonal directions. The pads on both sides are geometrically non-overlapping to minimize capacitive coupling, which is responsible for a non-localized component in the output signals (e.g. see the "precursor signal" in a delay line readout (fig. 7 in [17]).

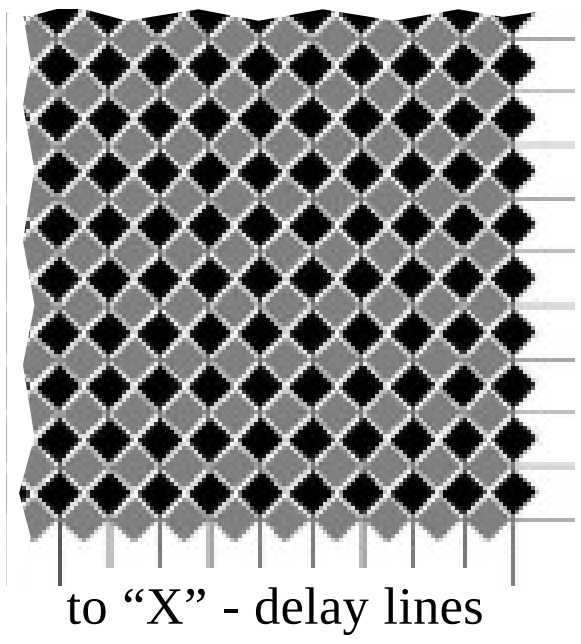

Fig. 2: Schematic view of a section of the RO electrode with the black (grey) pads in the front (back) side for recording of the "X" coordinates.

The position information is encoded by discrete LC-delay lines of 1,25 ns/mm mounted on the RO board using SMD. This is a simple and inexpensive technique, which can handle up to $10^6$ events /s, provided a capable electronic RO and data processing system is available. The presently available low cost system, based on the F1-TDC and ATMD board of ACAM electronics [18] can handle about $10^5$ $s^{-1}$ per module.

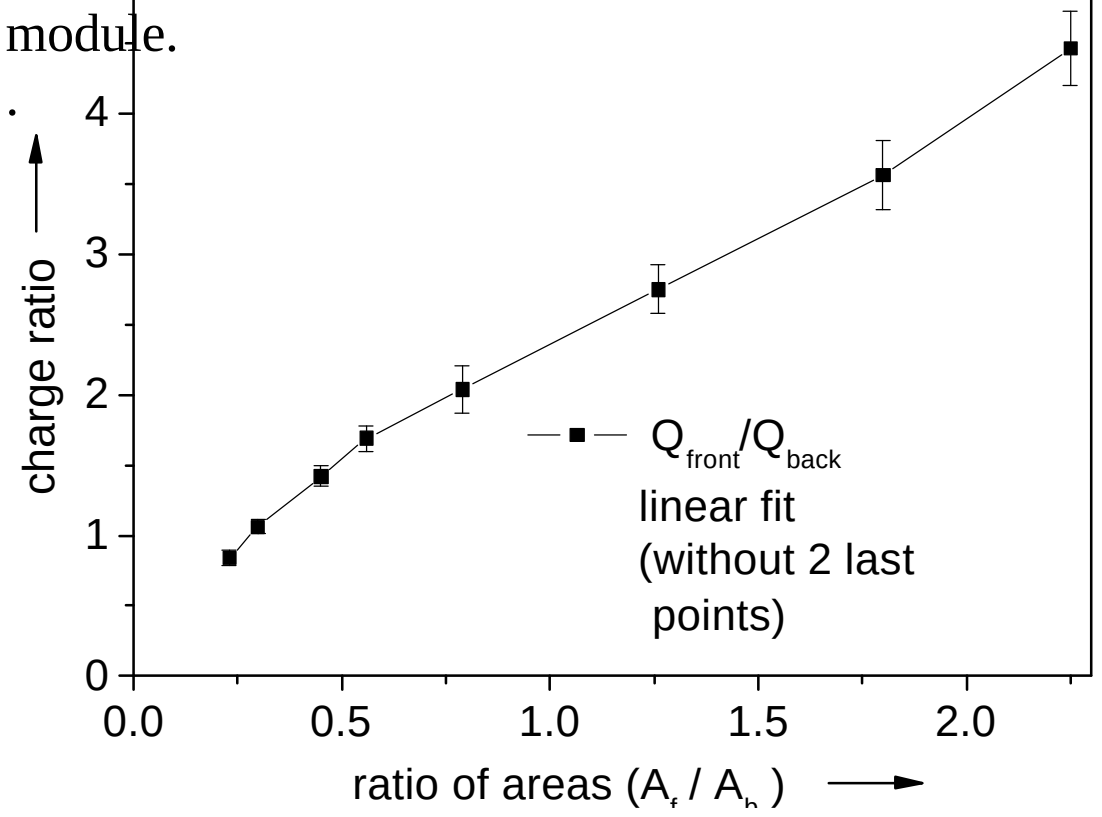

Fig 3: Charge ratio of fast signals from front and back side of RO board vs ratio or the areas covered by the pads on font ($A_f$) and back ($A_b$) side of the board

Significant efforts were made to optimize the geometry of the RO electrodes, namely the distance between the resistive anode and the RO electrode and the size of the pads. A full report and discussion of these measurements will be published in a separate paper. Here only some relevant results are summarized

The optimum geometry for the R/O electrode is the one with equal charge on front - ($Q_{front}$ )and back - side ($Q_{back}$), i.e. charge ratio $Q_{front}/Q_{back}$ = 1. Fig 3 shows the result of measurements with several electrodes, each with a different area ratios, covered by the pads on front ($A_f$) and backside($A_b$).

Another important factor is the radial charge spread. To obtain the highest position resolution without modulations in the position response by the periodic RO structure, the radial spread the induced charge should be close to the pitch of the RO structure. This pitch is the spacing between the pads, i.e. 2 mm. The radial width of the induced charge is determined mainly by the distance between the resistive anode and the RO board.

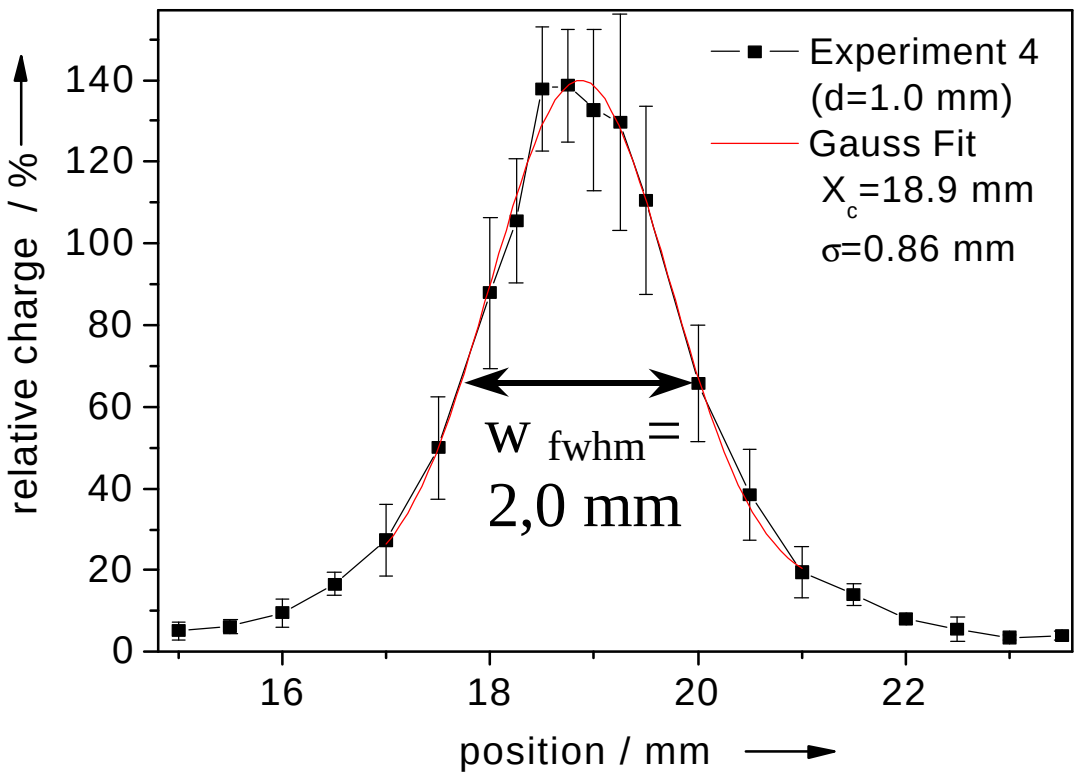

Fig. 4: Radial width of the induced charge signal at 1 mm distance (d) between resistive anode and RO electrode

We have measured this width for 2 distances, 2,6 and 1 mm, respectively, and found a radial spread of 5,1 and 2,0 mm. Fig 4 shows an example of the radial charge spread at a distance of 1 mm between the anode and the RO electrode. Based on these measurements the distance between the anode and the RO board was fixed at 1 mm.

## 3. Neutron Radiography with FANGAS

### 3.1. Energy resolution of FANGAS

FANGAS was developed for energy selective fast neutron radiography by measuring neutron TOF in pulsed beams. The method and setup for this experiment at PTB is described in [1]. A pulsed, broad spectrum neutron beam is produced by 13 MeV deuterium beam pulses of 1,5 ns width, hitting a thick Be target. The Be target is enclosed by a collimator which restricts the neutron beam to neutrons scattered into forward directions and defines a neutron field of appropriate size. The integral neutron flux at 3 m distance is $2 – 4 \times 10^5$ $s^{-1}$ $cm^{-2}$.

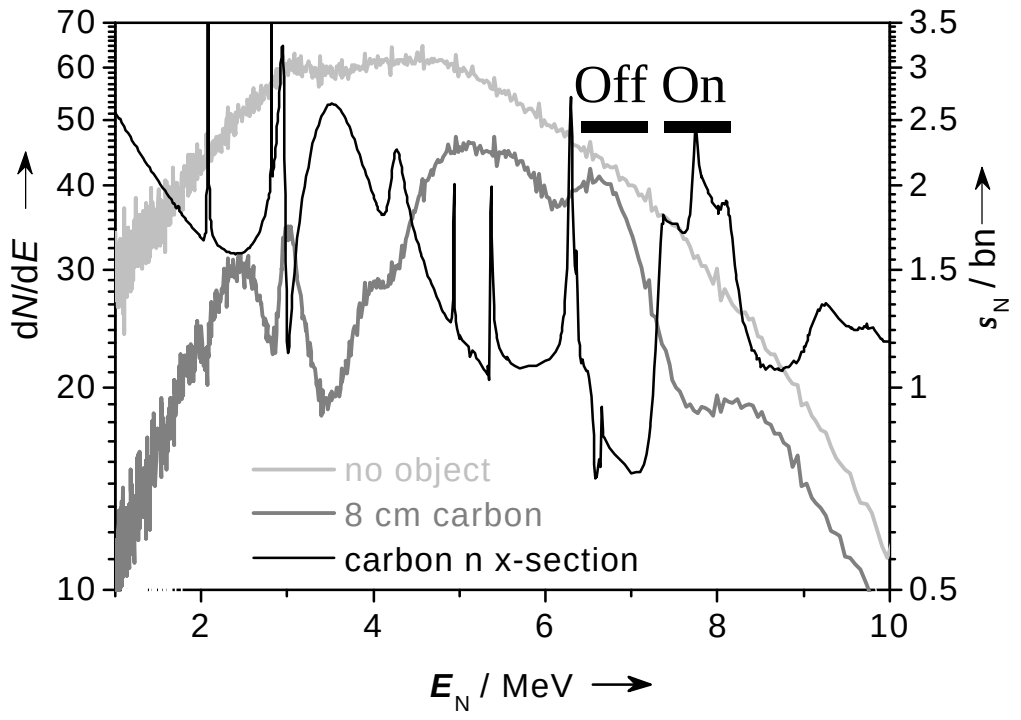

Fig 5: Energy spectra measured with FANGAS: light grey: without sample, dark grey: with 8 cm graphite sample at the collimator exit. The spectra are not corrected for the efficiency of the detector. The black curve shows the carbon cross section (scale to the left) The 2 bars marked “On” and “Off” show the energy windows used for the resonance imaging in sec. 3.3.

Fig. 5 (light grey curve) shows the neutron spectrum measured by a FANGAS module 3 m away from the target. The spectrum is not corrected for the detector efficiency and is derived directly from the TOF measurement. The dark grey curve shows the energy spectrum attenuated by an 8 cm thick graphite block which was inserted at the output of the collimator. The black curve shows the carbon cross section (scale at the right side). The structures of the C-cross section are reflected in the structures of the attenuated spectrum, showing dips in the yield where the cross section has peaks. For the test of the resonance imaging method (see chapter 3.3) the broad cross section structure between 6,5 and 7,3 MeV (OFF in fig. 5) and 7,4 and 8,4 MeV (ON in fig. 5) were chosen.

From the two energy spectra in fig. 5 the energy dependent transmission through the 8 cm graphite block is calculated and compared with the theoretical transmission probability. In the calculation the contribution of scattered neutrons from the sample can be neglected because of the distance of 2,6 m between C block and detector. Below 5 MeV the calculation is in good accordance with the experiment – apart of the sharp resonances, which of course cannot be resolved by this method.

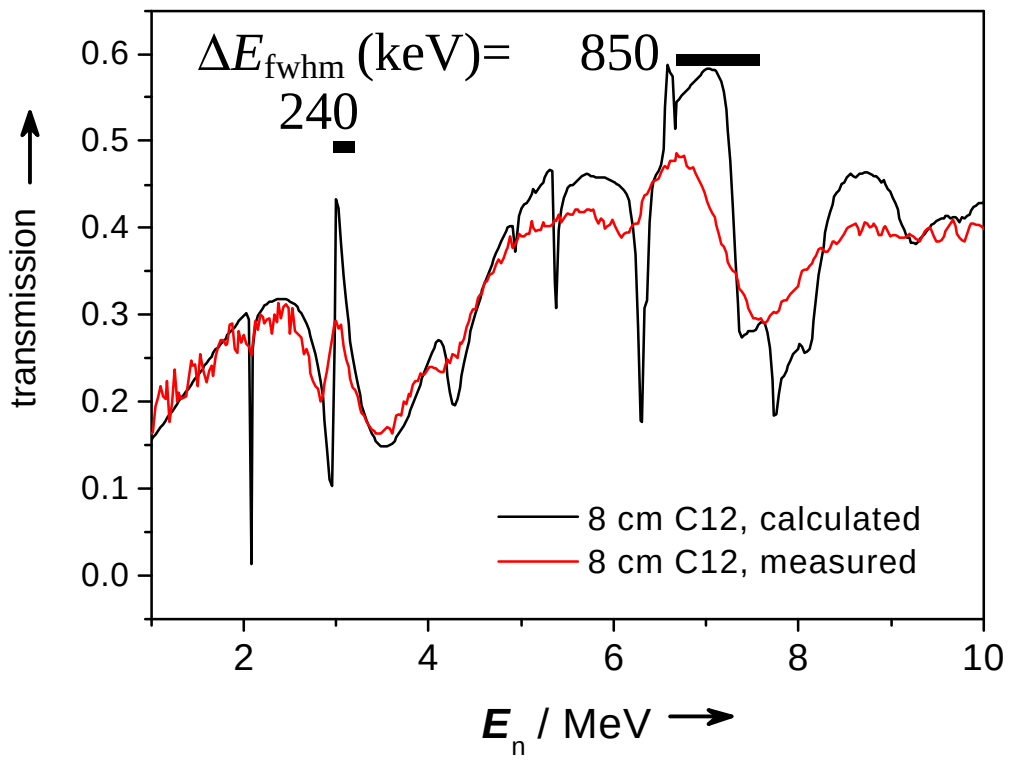

Fig 6: Theoretical and measured neutron transmission through a 8 cm thick graphite sample. The bars at 3 and 8 MeV show the energy resolution of the system at these energies.

However, in the region between 6 and 9 MeV where carbon has broad resonances, the deviation is significant. The reason is the rather poor time resolution of the FANGAS with $\Delta t_{fwhm}$ = 5 ns. At a neutron flight path of 3 m this corresponds to a energy resolution of $\Delta E_{fwhm}$ = 240 and 850 keV (fwhm) at $E_n$ = 3 and 7 MeV respectively (see black bars in Fig 6).

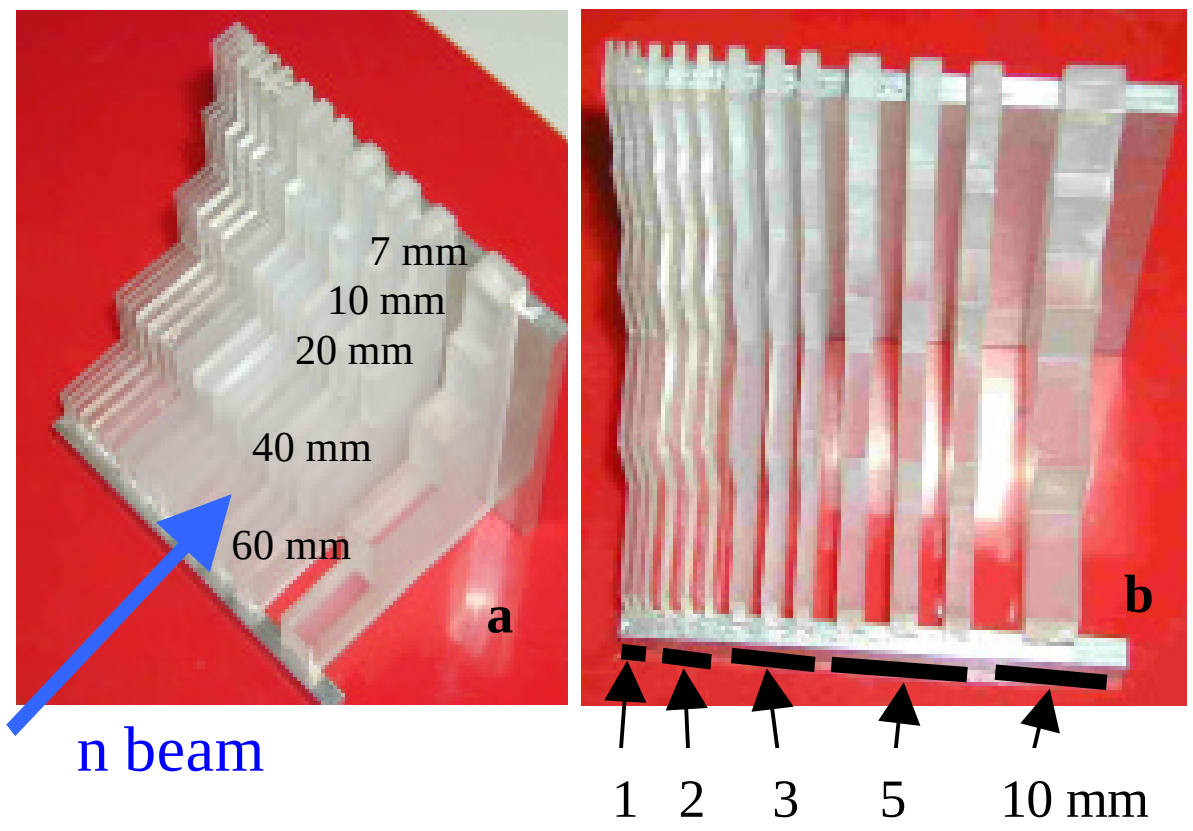

Fig 7: Step wedge mask made of polyvinyltoluene for the neutron radiography experiment. The width of the leaves varies from 1 – 10 mm (b) and the attenuation length in beam direction from 7 – 60 mm (see a).

An improvement of the time resolution to about 2 ns, which is close to the limit of the accelerator time resolution, is desirable and should be possible with this kind detector.

### 3.2. Position Resolution of FANGAS

The position resolution of FANGAS was measured with a step-wedge mask made of polyvinyltoluene (NE102 plastic scintillator material). The mask consists of a set of leaves with thickness ranging from 1 – 10 mm. Their length in beam direction ranges from 7 – 60 mm (see fig. 7). Fig. 8 shows the image of the step wedge in the neutron beam, obtained with the full neutron spectrum.

To calculate the position resolution, for 2 step sizes, 40 and 60 mm, the projections to the horizontal axis are calculated. (see white boxes fig. 8 and fig 9 as example for a projection of the 60 mm step). From these histograms the position resolution of the detector is calculated, assuming that the point spread function (PSF) of the detector is of Gaussian shape.

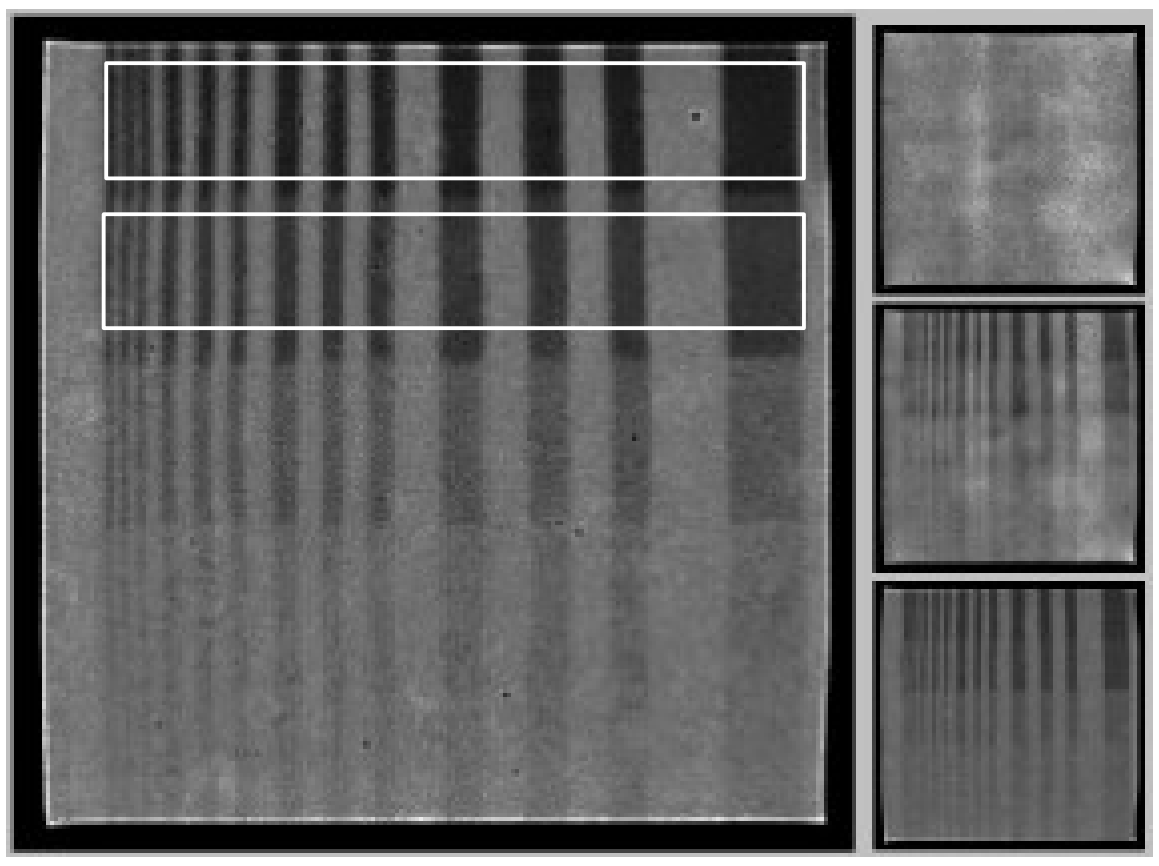

Fig 8 Neutron image of the step wedge mask. The small images at the right are: top:flat field, center: raw image, bottom: processed image (flat-field correction). The white boxes mark two areas which are analysed to derive the position resolution (see fig. 9)

The image is then a convolution of the rectangular mask and the PSF of the detector. The width of the Gaussian PSF is determined using Bayesian parameter estimation [19]. For the calculation, the program WinBUGS [20] was used.

From the histograms of the 60 and 40 mm long leaves, the width of the PSF was determined to be $\sigma_{60}$ = 0,45 (±0,03) and $\sigma_{40}$ = 0,43 ±0,03 mm (fwhm = 1,00 / 1,06 mm). Fig. 9 shows the measured (dots) and the MC simulated data for the 60 mm leaves.

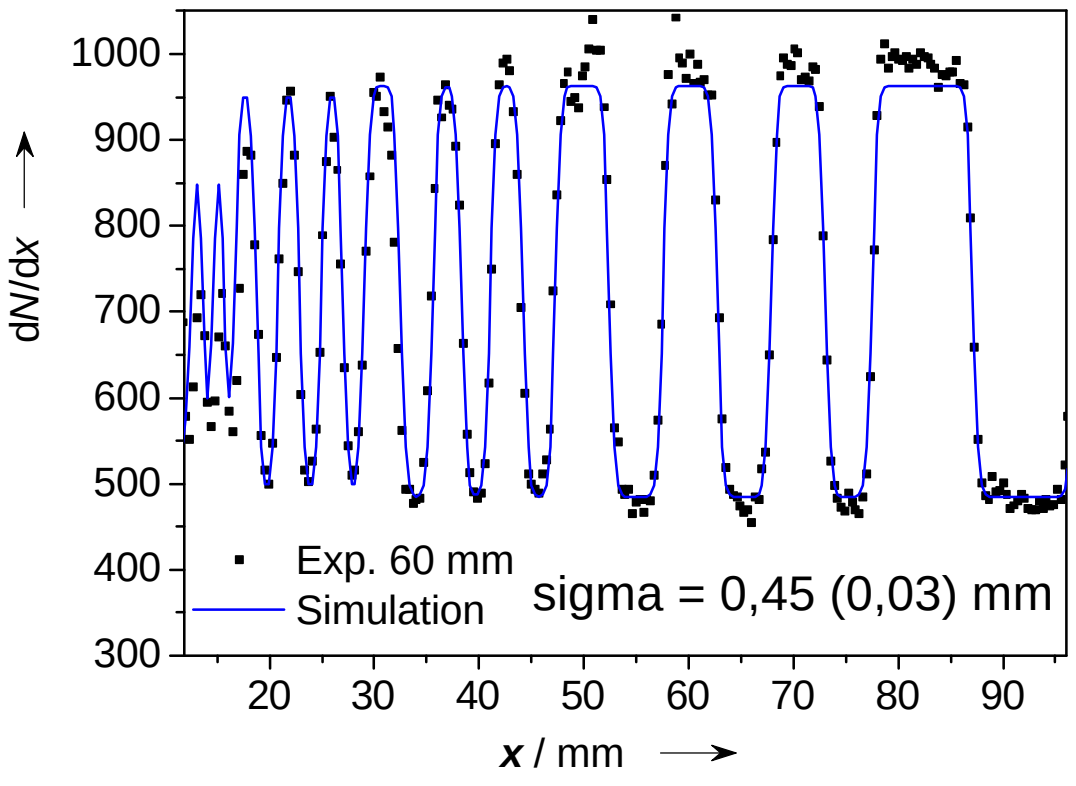

Fig. 9: Intensity modulation of 60 mm long NE102 leaves (dots) and fit by a rectangular/Gaussian folding model using Bayesian parameter estimation [19].

### 3.3. Resonance Imaging with Carbon

The resonance imaging method was tested on a composite sample made of carbon rods and a steel wrench (see fig. 10a,b). The setup of detector and sample is shown in 10c. Fig. 10d shows the integral (i.e. all energies) neutron image after flat field correction. In the energy range between 6 and 8 MeV the neutron cross section for steel is without significant variations [21] while carbon shows broad structures (see fig. 5).

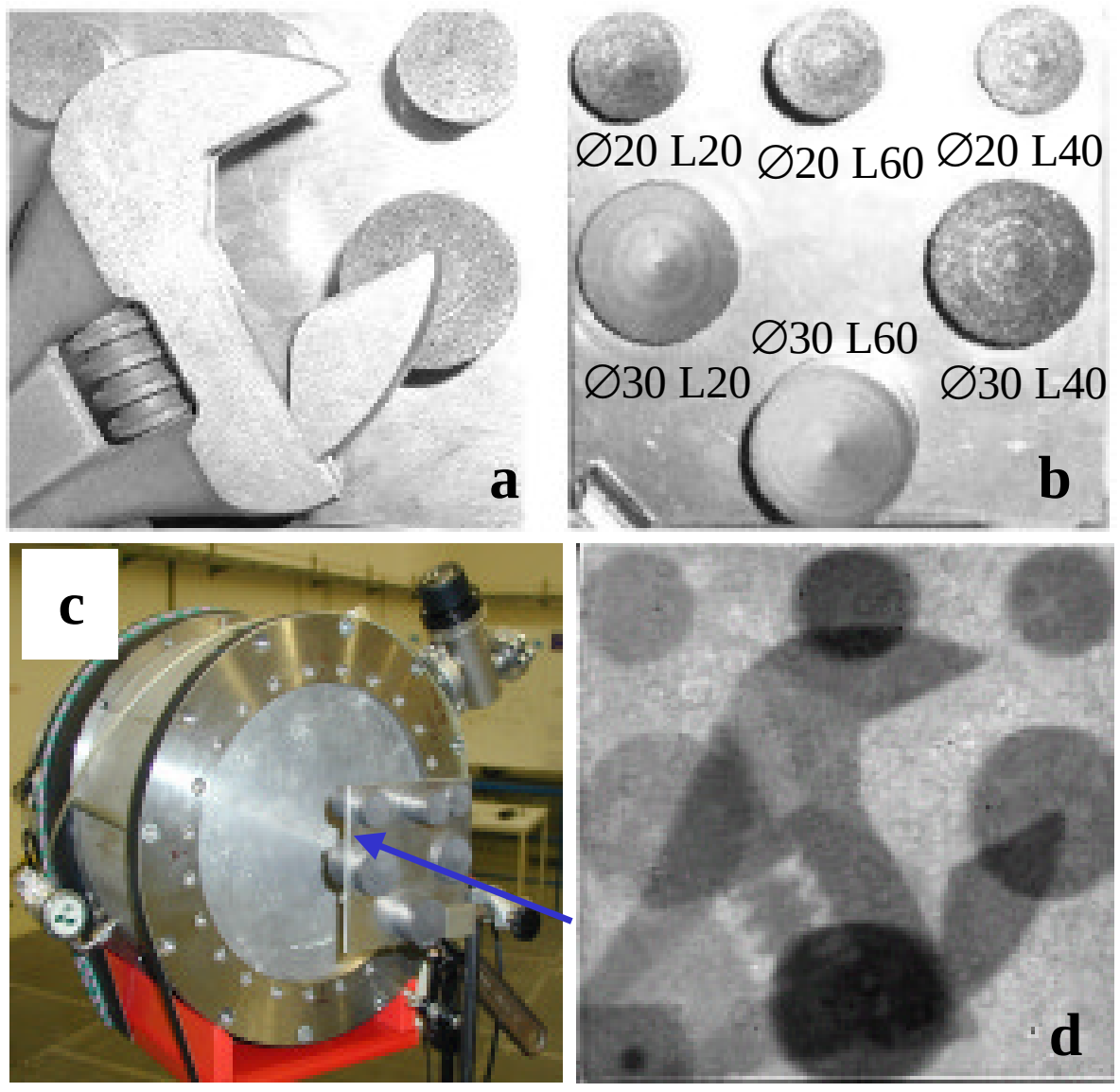

Fig.10: Imaging of carbon rods and a steel wrench with fast neutrons: . a,b) photos of sample, which consists of a steel wrench and carbon various rods. c) detector and sample in the neutron beam and d) neutron transmission image in a broad spectrum neutron beam.

Two images are selected, measured at energies marked as ON and OFF in fig. 5 and fig 11a. Dividing both images pixel-by-pixel is expected to result in a new image where the steel is eliminated and only the carbon samples remain visible.

For this measurement each registered neutron event was stored in list-mode, thus maintaining the full correlation between position and TOF. The ON- and OFF-resonance images were then reconstructed off-line according to the appropriate TOF windows (see fig. 11a), corresponding to the energy windows in fig. 7. The images with condition ON and OFF resonance are shown in fig 11. The resulting image, after dividing ON- and OFF-resonance images pixel-by-pixel is shown in fig.11R. As expected, the steel wrench disappears in the R image; however, also the thinner carbon samples cannot be resolved anymore. This poor contrast in the ratio image is not only due to the limited counting statistics in the raw images but mainly due to the poor time resolution (see fig. 5) which reduces the effect of the pronounced difference in neutron transmission between ON and OFF resonance.

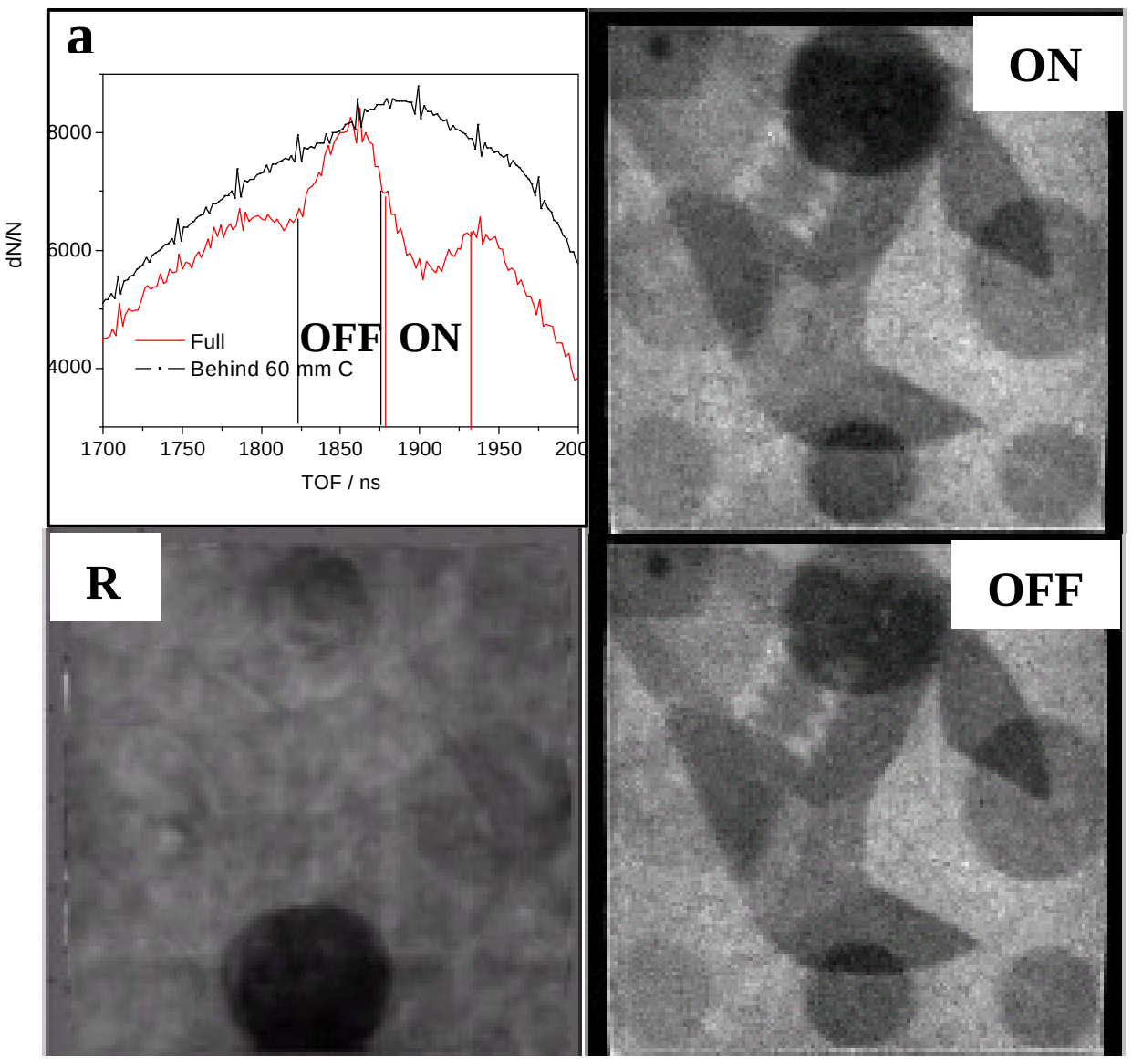

Fig. 11: Resonance imaging of mask of fig. 10.
a) TOF spectrum with windows at ON and OFF resonance. Pictures ON and OFF show the radiographs for the selected energy (TOF) windows, R is the pixel by pixel divided image ON / OFF.

## 4. Conclusion and Outlook

We have presented a pulse-counting neutron imaging detector with TOF capability and demonstrated its performance in a fast neutron resonance radiography experiment. At present we are developing a module with an active area of 30x30 $cm^2$ based on established GEM techniques. This detector will be the prototype for a cascade of 25 identical modules, stacked along the beam axis in order to achieve a total detection efficiency of about 5 %.

V. D. acknowledges the support of the Minerva foundation. A.B. is the W. P. Reuther Professor of Research in the Peaceful use of Atomic Energy

## References

[1] V. Dangendorf, G. Laczko, C. Kersten, O. Jagutzki, U. Spillmann, Fast Neutron Resonance Radiography in a Pulsed Neutron Beam, submitted for publication in IEEE Trans. of Nucl. Science (pp: http://arxiv.org/abs/nucl-ex/0301001)

[2] V. Dangendorf, A. Breskin, R. Chechik, G. Feldman, M.B. Goldberg, O. Jagutzki, C. Kersten, G. Laczko, I. Mor, U. Spillman, D. Vartsky, Detectors for Energy-Resolved Fast Neutron Imaging, submitted for publication in Nucl. Instr. and Meth. (2004), pp: http://arxiv.org/abs/nucl-ex/0403051

[3] V. Dangendorf, A. Demian, H. Friedrich, V. Wagner, A.Breskin, R. Chechik, S. Gibrekhterman, A. Akkerman, Nucl. Inst. and Meth. A350 (1994) 503

[4] A. Breskin, R. Chechik, A. Gibrekhterman, A. Akkerman, V. Dangendorf, A. Demian, Proc. SPIE, Vol 2339 (1995) 281

[5] A. Bäuning-Demian, H. Schmidt-Böcking, V. Dangendorf, H. Friedrich, A. Breskin, R. Chechik, A. Gibrekhterman: Proc. SPIE, Vol 2867 (1997) 562

[6] H. Friedrich, V. Dangendorf, A. Bräuning-Demian, App. Phys. A 74 (2002) 124

[7] F. Sauli, Nucl.Instr. and Meth. A386 (1997) 531

[8] GEANT 3.21, Detector Description and Simulation Tool, CERN Program Library Long Writeup W5013, CERN, Geneva, 1993

[9] D. Vartsky et al, Time-Resolved Fast Neutron Imaging: Simulation of System Performance, this proceedings

[10] J. F. Ziegler, J. P. Biersack and U. Littmark, SRIM - The Stopping and Range of Ions in Matter, http://www.srim.org

[11] G.P. Guedes, A. Breskin, R. Chechik, D. Mörmann, Nucl.Instr. and Meth .A497 (2003) 305

[12] O. Jagutzki, J.S. Lapington, L.B.C Worth, U. Spillmann, V. Mer-gel, H. Schmidt-Böcking, Nucl. Instr.and Meth. A477 (2002) 256

[13] M.S.Dixit, J.Dubeau, J.-P. Martin C. K. Sachs, Nucl. Instr. and Meth. A518 (2004) 721

[14] R. Baur, A. Drees, P. Fischer, P. Glässel, D. Irmscher, A. Pfeiffer, A. Schön, Ch. Schwick, H.J.Specht, S. Tapprogge, Nucl. Instr. and Meth. A355 (1995) 329

[15] G. Battistoni et al, Resistive Cathode Transparency, Nucl. Instr. and Meth. 202(1982)459

[16] Meir Shoa, Weizmann Institute of Science/Israel, Resistive anodes for the Atlas TGC, private communication 2003, see also S. Tanaka, Techniques developed for the ATLAS TTGC, H.ATL- MUON- 2004- 015, http://doc.cern.ch//archive/electronic/cern/ others/ atlnot/Note/muon/muon-2004-015.pdf

[17] G.P.Guedes, A.Breskin, R. Chechik, D.Vartsky, D.Bar, A.F. Barbosa, P.R.B.Marinho, Nucl.Instr. and Meth. A513 (2003) 473

[18] ACAM-Messelectronic GmbH, www.acam.de

[19] D. S. Sivia, Data Analysis - A Bayesian Tutorial, Clarendon Press, Oxford, 1996

[20] D. Spiegelhalter, A. Thomas, N. Best, W. Gilks, D. Lunn, (1994, 2003). BUGS: Bayesian inference using Gibbs sampling. MRC Biostatistics; Cambridge, England, www.mrc-su.cam.ac.uk/bugs/

[21] Nuclear Data Evaluation Lab., Korea Atomic Energy Research Institute, ENDFPLOT 2.0, http://atom.kaeri.re.kr/